\documentclass[11pt,epsf]{article}
\usepackage{amsmath}
\usepackage{amsfonts}
\usepackage{amssymb}
\usepackage{graphicx}
\usepackage{color}
\usepackage{comment}
\newtheorem{theorem}{Theorem}

\newcommand{\hH}{\hat{H}}

\newcommand{\tx}{\tilde{x}}
\newcommand{\ty}{\tilde{y}}
\newcommand {\dfn} {\stackrel{\Delta} {=}}

\newcommand{\lea}{\stackrel{\mbox{\tiny (a)}}{\le}}
\newcommand{\leb}{\stackrel{\mbox{\tiny (b)}}{\le}}
\newcommand{\lec}{\stackrel{\mbox{\tiny (c)}}{\le}}
\newcommand{\led}{\stackrel{\mbox{\tiny (d)}}{\le}}

\newcommand {\bs} {\mbox{\boldmath $s$}}

\newcommand {\bu} {\mbox{\boldmath $u$}}
\newcommand {\bv} {\mbox{\boldmath $v$}}

\newcommand {\bx} {\mbox{\boldmath $x$}}
\newcommand {\by} {\mbox{\boldmath $y$}}
\newcommand {\bz} {\mbox{\boldmath $z$}}

\newcommand{\calA}{{\cal A}}
\newcommand{\calB}{{\cal B}}

\newcommand{\calE}{{\cal E}}

\newcommand{\calI}{{\cal I}}

\newcommand{\calR}{{\cal R}}
\newcommand{\calS}{{\cal S}}
\newcommand{\calT}{{\cal T}}
\newcommand{\calU}{{\cal U}}
\newcommand{\calV}{{\cal V}}

\newcommand{\calX}{{\cal X}}
\newcommand{\calY}{{\cal Y}}
\newcommand{\calZ}{{\cal Z}}

\begin{document}
\thispagestyle{empty}
\title{Universal Slepian-Wolf Coding for Individual Sequences}
\author{Neri Merhav}
\date{}
\maketitle

\begin{center}
The Andrew \& Erna Viterbi Faculty of Electrical and Computer Engineering\\
Technion - Israel Institute of Technology \\
Technion City, Haifa 32000, ISRAEL \\
E--mail: {\tt merhav@ee.technion.ac.il}\\
\end{center}
\vspace{1.5\baselineskip}
\setlength{\baselineskip}{1.5\baselineskip}

\begin{abstract}
We establish a coding theorem and a matching converse theorem for separate
encodings and joint decoding of individual sequences using finite-state
machines. The achievable rate region is characterized in terms of the
Lempel-Ziv (LZ) complexities, the conditional LZ complexities and the joint LZ
complexity of the two source sequences. An important feature that is needed to
this end, which may be interesting on its own right, is a certain asymptotic
form of a chain rule for LZ
complexities, which we establish in this work.
The main emphasis in the achievability
scheme is on the universal decoder and its properties. 
We then show that the achievable rate region is
universally attainable by a modified version of Draper's universal incremental
Slepian-Wolf (SW) coding scheme, provided that there exists a low-rate
reliable feedback link.
\end{abstract}


\clearpage
\section{Introduction}
\label{intro}

The renowned Slepian-Wolf (SW) source coding theorem, first introduced by
Slepian and Wolf in 1973 \cite{SW73}, unveils a captivating revelation in the
realm of (almost) lossless, fixed-rate compression for memoryless sources in
the presence of side information. Remarkably, the theorem establishes that the
conditional entropy of the source, given the side information, can be achieved
through random binning even if the side information
is exclusively available at the decoder, without a necessity for its presence
at the encoder.
Expanding the horizons of the Slepian-Wolf setting, the theorem is
instrumental in characterizing the rate region associated with separate
encodings and joint decoding of two correlated memoryless sources. In such
scenarios, each coding rate is independently lower bounded by the
corresponding conditional entropy, while the rate sum finds its lower bound in
the joint entropy. It is imperative to note that in both settings, the joint
distribution of the two correlated sources is assumed to be known.

In \cite[Problem 13.6, p.\ 267]{CK11}, Csisz\'ar and K\"orner considered the
case where the joint distribution of the correlated sources is unknown. In
this case, the
encoders continue to use random binning as before, but the optimal maximum
a-posteriori (MAP) decoder is
replaced by a universal decoder that seeks a
pair of sequences (across the given bins) with minimum joint empirical
entropy (see also \cite{me16}, as well as references therein,
for a wider setup of universal source-channel coding and decoding for
finite-state sources and finite-state channels with side information).
As indicated by Draper \cite{Draper04}, the obvious weakness of Csisz\'ar and
K\"orner's universal scheme is that one must commit to fixed coding rates although the source
statistics are unknown, and so, there is no mechanism that could adapt the coding
rates to the corresponding entropies. Motivated by the will to circumvent this problem, and inspired
of earlier works by Shulman \cite{Shulman03} and Shulman
and Feder \cite{SF02}, Draper
proposed a universal, incremental, variable-rate coding scheme that can be
implemented provided that a low rate, reliable feedback link is available.

Another perspective of universal source coding is associated with the individual-sequence setting and
finite-state machines, as explored by Ziv and Lempel in their celebrated work
\cite{ZL78} among some other papers. Indeed, in \cite{Ziv84}, Ziv studied a scenario of fixed-rate
coding with side information where both the source sequence and the side
information sequence are individual (deterministic) sequences and where the
encoder and decoder are both implementable by finite-state machines. The main
finding in \cite{Ziv84} is in establishing and characterizing a notion of fixed-rate conditional
complexity as the minimum, almost-lossless compression rate of a source sequence given a side information
sequence, and similarly as in classical SW coding, the availability of the
side information at the encoder is not necessary in order to achieve this
conditional complexity. A year later, in \cite{Ziv85}, a variable-rate version of the
conditional Lempel-Ziv (LZ) complexity was proposed in the completely different context of serving as a universal
channel decoding metric for unknown finite-state channels. The utility of
this complexity measure in the context of source coding with side information
was given further attention 
later in \cite{me00} and \cite{UK03}, but in these works, it was assumed that
the side information is available at both ends.

In this work, we consider the framework of SW coding
for individual sequences using finite-state encoders. We begin by
establishing a coding theorem and a matching converse that together
characterize the achievable rate region. Our communication system model is different from that of
\cite{Ziv84} in several aspects: (i) We consider separate encodings and joint
decoding of two individual sequences, as opposed to the narrower problem of encoding a single sequence
with the other sequence serving as decoder side information; (ii) Our model
for the converse theorem allows variable-rate, finite-state encoding and
arbitrary decoding, as opposed to fixed-rate, finite-state encoding and
finite-state decoding of \cite{Ziv84}; (iii) The relation to the variable-rate coding model in
\cite{ZL78} is more apparent; (iv) We establish the variable-rate conditional
LZ complexity as the fundamental limit even when the side information is
available at the decoder only, and not only when it is available at both ends.

The characterization of the achievable rate region raises an issue which may
be interesting on its own right: Recall that in the classical regime of
two correlated discrete memoryless sources, $X$ and $Y$, with joint entropy
$H(X,Y)$ and conditional entropies
$H(X|Y)$ and $H(Y|X)$, the
achievable rate region is given by $\{(R_{\mbox{\tiny x}},R_{\mbox{\tiny
y}}):~R_{\mbox{\tiny x}}\ge H(X|Y),~R_{\mbox{\tiny y}}\ge H(Y|X), R_{\mbox{\tiny
x}}+R_{\mbox{\tiny y}}\ge H(X,Y)\}$, whose corner points are given by
$(H(X),H(Y|X))$ and
$(H(X|Y),H(Y))$, where the 
appearance of unconditional marginal entropies, $H(X)$ and $H(Y)$, follows from the chain rule of the entropy,
$H(X,Y)=H(X)+H(Y|X)=H(Y)+H(X|Y)$. In the individual-sequence scenario,
as we shall see, the achievable region is similar except that
$H(X|Y)$, $H(Y|X)$ and $H(X,Y)$ are replaced by the corresponding conditional LZ complexities 
of the two source sequences and their joint complexity, respectively. However, there is no apparent exact corresponding chain rule
that analogously decomposes the joint LZ complexity of two source sequences,
$x^n=(x_1,\ldots,x_n)$ and
$y^n=(y_1,\ldots,y_n)$ as the LZ complexity of $x^n$ plus the conditional LZ complexity of
$y^n$ given $x^n$ or vice versa. Nonetheless, we will show that at least in a certain
asymptotic sense, such as chain rule between the LZ complexities actually applies. This
will be instrumental for nailing down the characterization of the corner
points of the achievable region in an appealing manner.

In the second part of the paper, we propose a modification of Draper's
incremental scheme that is suitable to individual sequences along with their
LZ complexities. This will be possible by drawing simple analogies between the
various ingredients in Draper's scheme to their corresponding analogues in our
individual-sequence setting.

The outline of the remaining part of the paper is as follows. 
In Section \ref{nfob}, we establish notation conventions, formulate the
problem model, define the objectives of this work, and provide some background.
In Section \ref{codingtheorem}, we assert and discuss the main coding theorem
(Theorem \ref{thm1})
and also establish the asymptotic ``chain rule'' of the LZ complexity (Theorem
\ref{thm2}). Finally, in Section \ref{incremental}, we describe and analyze
the universal incremental coding scheme that adjusts the coding rates
dynamically. Lengthy proofs are deferred to appendices.

\section{Notation, Formulation, Objectives and Background}
\label{nfob}

\subsection{Notation}
\label{notation}

Throughout the paper, random variables will be denoted by capital
letters, specific values they may take will be denoted by the
corresponding lower case letters, and their alphabets
will be denoted by calligraphic letters. Random
vectors, their realizations and their alphabets will be denoted,
respectively, by capital letters, the corresponding lower case letters, and
the corresponding calligraphic letters, all
superscripted by their dimension.
For example, the random vector $X^n=(X_1,\ldots,X_n)$, ($n$ --
positive integer) may take a specific vector value $x^n=(x_1,\ldots,x_n)$
in $\calX^n$, the $n$--th order Cartesian power of the single-letter alphabet $\calX$,
which will be assumed to have a finite cardinality, $\alpha$. The notation $x_i^j$, for
$i<j$, will be used to designate the substring $(x_i,x_{i+1},\ldots,x_j)$.
For $i=1$, the subscript $i$ will be omitted, just like in the notation $x^n$.
Infinite sequences will be
denoted using the bold face font, for example, $\bx$ will designate the
sequence $(x_1,x_2,\ldots)$. Similar conventions will apply to other vectors
and sequences, such as $y^n\in \calY^n$, $y_i^j$, and $\by$. The single-letter
alphabet $\calY$ will also be assumed to be finite and its cardinality will be denoted by
$\beta$. The probability of an event $\calA$ will be denoted by $\mbox{Pr}\{\calA\}$.
Entropies will be denoted using the customary information-theoretic notation,
like $H(X)$, $H(Y)$,
$H(X,Y)$, $H(X|Y)$, etc., and the same holds for random vectors. 
The indicator function
of an event $\calA$ will be denoted by $\calI\{A\}$. The notation $[x]_+$
will stand for $\max\{0,x\}$. The cardinality of a finite set $\calA$ will be
denoted by $|\calA|$.

For a given positive integer $n$ and a given $\ell$ that divides $n$,
the empirical distribution of non-overlapping $\ell$-blocks associated with a
vector pair $(x^n,y^n)\in\calX^n\times\calY^n$, which will be
denoted by $\hat{P}$, is the set of relative frequencies
\begin{equation}
\hat{P}(x^\ell,y^\ell)=\frac{\ell}{n}\sum_{i=0}^{n/\ell-1}
\calI\{(x_{i\ell+1}^{i\ell+\ell},y_{i\ell+1}^{i\ell+\ell})=(x^\ell,y^\ell)\},~~~~(x^\ell,y^\ell)\in\calX^\ell\times\calY^\ell.
\end{equation}
Hereafter, $X^\ell$ and $Y^\ell$ will denote auxiliary random vectors of dimension $\ell$,
jointly distributed according to $\hat{P}$. Accordingly, we will denote
by $\hH(X^\ell)$, $\hH(Y^\ell)$, $\hH(X^\ell,Y^\ell)$, $\hH(X^\ell|Y^\ell)$,
etc., the various entropies and conditional entropies associated with $(X^\ell,Y^\ell)$.
In the sequel, we will also divide the pair of $n$-vectors $(x^n,y^n)$ into $n/k$
non-overlapping segments, each of length $k$ (where $k$ divides $n$) and
within each such segment, $(x_{ik+1}^{ik+k},y_{ik+1}^{ik+k})$, $i=0,1,\ldots,n/k-1$,
we define the empirical distribution of non-overlapping $\ell$-blocks, $\hat{P}_i$ (assuming
that $k$ is divisible by $\ell$), and denote
the corresponding auxiliary random $\ell$-vectors by $X_i^\ell$ and $Y_i^\ell$,
respectively. Clearly,
$\hat{P}(x^\ell,y^\ell)=\frac{k}{n}\sum_{i=0}^{n/k-1}\hat{P}_i(x^\ell,y^\ell)$.

\subsection{Formulation}
\label{formulation}

Let $\bx=(x_1,x_2,\ldots)$ and
$\by=(y_1,y_2,\ldots)$ 
be two individual sequences whose single-letter alphabets, $\calX$ and
$\calY$, have finite
cardinalities, $\alpha$ and $\beta$, respectively. Both sequences are to be
compressed almost losslessly by separate encoders and jointly decompressed by
a central decoder. Each one of encoders is a finite-state encoder defined
similarly as in \cite{ZL78}, but with a small twist that allows some
arbitrarily small distortion (to make the model broad enough to include
Slepian-Wolf coding). Also, since the formulation of the setting in this
section is for the purpose of the converse bound, it is legitimate to broaden
the class of the encoders by allowing them to be ``genie-aided'' encoders,
i.e., letting each one of
them to have (sequential) access to the other source as side information. Of course, in the achievability
scheme, such an access will not be allowed. 

The encoder for $\bx$, henceforth referred to as the {\em x-encoder}, 
is defined by the set 
$$E_{\mbox{\tiny x}}=(\calS,
\calX,\calY,\calU,
g_{\mbox{\tiny x}},f_{\mbox{\tiny x}}),$$
where $\calS$ is
the set of states, $\calX$ is the input alphabet as mentioned,
$\calY$ is the side information alphabet,
$\calU$ is a finite set of binary output
strings, $g_{\mbox{\tiny x}}:\calS\times\calX\times\calY\to\calS$ is the next-state function,
and $f_{\mbox{\tiny x}}:\calS\times\calX\times\calY\to\calU$ is the
output function. 
The members of $\calU$ are allowed to be of
different lengths, including the empty word $\lambda$ of length zero.
When the input $\bx=(x_1,x_2,\ldots)$ feeds the encoder 
$E_{\mbox{\tiny x}}$, it produces an output sequence
$\bu=(u_1,u_2,\ldots)$, $u_i\in\calU$, $i=1,2,\ldots$, while traversing an
infinite sequence of states, $\bs=(s_1,s_2,\ldots)$, $s_i\in\calS$, in
accordance to
\begin{eqnarray}
u_i&=&f_{\mbox{\tiny x}}(s_i,x_i,y_i)\\
s_{i+1}&=&g_{\mbox{\tiny x}}(s_i,x_i,y_i),~~~~i=1,2,\ldots
\end{eqnarray}
where $s_1$ is assumed a fixed member of $\calS$. 
The functions $f_{\mbox{\tiny x}}$ and $g_{\mbox{\tiny x}}$ are allowed to be
``slowly time-varying'' in the sense that for some large positive integer $k$,
these functions are piece-wise constant over blocks of length $k$. In other
words, $f_{\mbox{\tiny x}}$ and $g_{\mbox{\tiny x}}$ are allowed to depend 
on the running time index $i$ via the quantity $\lfloor i/k\rfloor$. To avoid cumbersome
notation, however, we will not indicate this dependence explicitly in the mathematical
derivations. Eventually, the parameter $k$ will tend to infinity, which means
that the temporal variability is asymptotically slower than that of any
time-varying system.

The encoder for $\by$, henceforth referred to as the {\em y-encoder}, 
is defined exactly in the same manner, except 
that the roles of the two sources are swapped and accordingly,
the subscript ``x'' is replaced
by ``y'' in all places. Also, $\calS$ is replaced by $\calZ$, $\calU$ is replaced by
$\calV$, and accordingly, $\bs$, $s_i$, $\bu$ and $u_i$ are substituted by
$\bz$, $z_i$, $\bv$ and $v_i$, respectively. Without an essential loss of generality, the number of states of both
$E_{\mbox{\tiny x}}$ and 
$E_{\mbox{\tiny y}}$, namely $|\calS|$ and $|\calZ|$, will be assumed the
same, and both will be denoted by $q$.

As in \cite{ZL78}, we adopt the extended notation $f_{\mbox{\tiny
x}}(s_i,x_i^j,y_i^j)$
and $g_{\mbox{\tiny x}}(s_i,x_i^j,y_i^j)$ to denote the output segment $u_i^j$ and
final state $s_j$ resulting when the input string $x_i^j$ feeds
$E_{\mbox{\tiny x}}$, which is at state $s_i$ at time $i$.
Likewise, $g_{\mbox{\tiny y}}(z_i,y_i^j,x_i^j)$ and
$f_{\mbox{\tiny y}}(z_i,y_i^j,x_i^j)$ play analogous roles
for the y-encoder.

Let $\epsilon\in(0,1)$ be a given arbitrarily small number.
We assume that $(E_{\mbox{\tiny x}},E_{\mbox{\tiny y}})$ 
together with their joint decoder $D$ form
an {\em $\epsilon$-lossy system}, which is defined as follows: For every
$(z_1,s_1)\in\calZ\times\calS$ and all
sufficiently large $\ell$, the six-tuple $(z_1,s_1,f_{\mbox{\tiny
x}}(s_1,x^\ell,y^\ell),f_{\mbox{\tiny y}}(z_1,y^\ell,x^\ell),g_{\mbox{\tiny
x}}(s_1,x^\ell,y^\ell),g_{\mbox{\tiny y}}(z_1,y^\ell,x^\ell))$ 
uniquely determines a decoder
output string pair
$(\hat{x}^\ell,\hat{y}^\ell)$, whose normalized Hamming distance
from $(x^\ell,y^\ell)$ does not exceed $\epsilon$, namely,
$\frac{1}{\ell}\sum_{i=1}^\ell\calI\{(\hat{x}_i,\hat{y}_i)\ne(x_i,y_i)\}\le\epsilon$.
In addition, the quadruplet $(y^\ell,s_1,f_{\mbox{\tiny
x}}(s_1,x^\ell,y^\ell),g_{\mbox{\tiny x}}(s_1,x^\ell,y^\ell))$ determines $\hat{x}^\ell$
within normalized Hamming distance $\epsilon$ away from $x^\ell$, and vice
versa: $(x^\ell,z_1,f_{\mbox{\tiny
y}}(z_1,y^\ell,x^\ell),g_{\mbox{\tiny y}}(z_1,y^\ell,x^\ell))$ yields $\hat{y}^\ell$,
whose normalized Hamming distance from $y^\ell$ is at most $\epsilon$.

Let $\calE(q,\epsilon)$ denote the class of all
$\epsilon$-lossy pairs of finite-state encoders, $(E_{\mbox{\tiny x}},E_{\mbox{\tiny y}})$,
with no more than $q$ states each. The total number of combinations of states
of $E_{\mbox{\tiny x}}$ with states of $E_{\mbox{\tiny y}}$ is therefore no more than
$q^2$.

\subsection{Objectives}
\label{objectives}

Given the vector pair $(x^n,y^n)$ formed by the first $n$ symbol pairs of
$(\bx,\by)$, let $u^n=(u_1,\ldots,u_n)=f_{\mbox{\tiny x}}(s_1,x^n,y^n)$ denote
the output of $E_{\mbox{\tiny x}}$. We define the compression ratio of $x^n$ by
$E_{\mbox{\tiny x}}$ as
\begin{equation}
\rho_{E_{\mbox{\tiny x}}}(x^n)=\frac{L(u^n)}{n},
\end{equation}
where $L(u^n)=\sum_{i=1}^nl(u_i)$, $l(u_i)$ being the length (in bits) of
$u_i$, and where it should be kept in mind that for the empty string
$\lambda$, we set
$l(\lambda)=0$. Likewise, for $v^n=(v_1,\ldots,v_n)=f_{\mbox{\tiny
y}}(z_1,y^n,x^n)$,
\begin{equation}
\rho_{E_{\mbox{\tiny y}}}(y^n)=\frac{L(v^n)}{n},
\end{equation}
with $L(v^n)=\sum_{i=1}^nl(v_i)$ and with $l(v_i)$ denoting the length of
$v_i$. 

For a given encoder pair $(E_{\mbox{\tiny x}},E_{\mbox{\tiny
y}})\in\calE(q,\epsilon)$, let
\begin{equation}
\calR_{E_{\mbox{\tiny x}},E_{\mbox{\tiny y}}}(\bx,\by)=\left\{(R_{\mbox{\tiny
x}},R_{\mbox{\tiny y}}):~R_{\mbox{\tiny
x}}\ge\limsup_{n\to\infty}\rho_{E_{\mbox{\tiny x}}}(x^n),~
R_{\mbox{\tiny
y}}\ge\limsup_{n\to\infty}\rho_{E_{\mbox{\tiny y}}}(y^n)\right\}.
\end{equation}
Next, define
\begin{equation}
\calR_{q,\epsilon}(\bx,\by)=\bigcup_{(E_{\mbox{\tiny x}},E_{\mbox{\tiny
y}})\in\calE(q,\epsilon)}\calR_{E_{\mbox{\tiny x}},E_{\mbox{\tiny
y}}}(\bx,\by),
\end{equation}
\begin{equation}
\calR_\epsilon(\bx,\by)=\bigcup_{q\ge 1}\calR_{q,\epsilon}(\bx,\by),
\end{equation}
and finally,
\begin{equation}
\calR(\bx,\by)=\bigcap_{\epsilon>0}\calR_\epsilon(\bx,\by).
\end{equation}
These definitions are essentially the two-dimensional counterparts of the
$s$-state compressibility of a single source vector $x^n$ \cite[eq.\ (2)]{ZL78}, the asymptotic $s$-state
compressibility of $\bx$ \cite[eq.\ (3)]{ZL78},
and the asymptotic finite-state compressibility of $\bx$ \cite[eq.\
(4)]{ZL78}.

Our first objective is to characterize the set of rate pairs, $\calR(\bx,\by)$.
Our second objective is to propose a universal, incremental variable-rate coding scheme that
asymptotically achieves $\calR(\bx,\by)$ with the aid of a low-rate,
reliable feedback channel, that allows adaptation of the coding rates to the
compressibilities of the two source sequences. We do that by a simple
modification of Draper's scheme for memoryless sources \cite{Draper04}.

\subsection{Background}
\label{background}

To support the exposition of both the converse theorem and the
achievability theorem, it is necessary to revisit key terms and details
related to the 1978 version of the LZ algorithm, also known as the LZ78
algorithm \cite{ZL78}.
The incremental parsing procedure of the LZ78 algorithm is a sequential parsing
process applied to the source vector $x^k$.
In this procedure, each
new phrase is the shortest string not encountered before as a parsed phrase,
except for the potential incompleteness of the last phrase. For instance, the
incremental parsing of the vector $x^{15}=\mbox{abbabaabbaaabaa}$ results in 
$\mbox{a,b,ba,baa,bb,aa,ab,aa}$. Let $c(x^k)$ denote the
number of phrases in $x^k$ resulting from the incremental parsing procedure
(in the above example, $c(x^{15})=8$).
Furthermore, let $LZ(x^k)$ denote the length of the LZ78 binary compressed
code for $x^k$. According to \cite[Theorem 2]{ZL78}, the following inequality
holds:
\begin{eqnarray}
\label{lz-clogc}
LZ(x^k)&\le&[c(x^k)+1]\log\{2\alpha[c(x^k)+1]\}\nonumber\\
&=&c(x^k)\log[c(x^k)+1]+c(x^k)\log(2\alpha)+\log\{2\alpha[c(x^k)+1]\}\nonumber\\
&=&c(x^k)\log c(x^k)+c(x^k)\log\left[1+\frac{1}{c(x^k)}\right]+
c(x^k)\log(2\alpha)+\log\{2\alpha[c(x^k)+1]\}\nonumber\\
&\le&c(x^k)\log c(x^k)+\log
e+\frac{k(\log \alpha)\log(2\alpha)}{(1-\varepsilon_k)\log
k}+\log[2\alpha(k+1)]\nonumber\\
&\dfn&c(x^k)\log c(x^n)+k\cdot\epsilon(k),
\end{eqnarray}
where we remind that $\alpha$ is the cardinality of $\calX$, and where
both $\varepsilon_k$ and $\epsilon(k)$ tends to zero as $k\to\infty$. 
In other words, the LZ code-length for $x^k$ is upper bounded by
an expression whose main term is $c(x^k)\log c(x^k)$. On the other hand,
$c(x^k)\log c(x^k)$ is also known to be the main term of a lower bound \cite[Theorem 1]{ZL78}
to the shortest code-length attainable by any information lossless finite-state encoder with no
more than $s$ states, provided that $\log(s^2)$ is very small compared to
$\log c(x^k)$. In view of these facts, we henceforth refer to $c(x^k)\log c(x^k)$ as the unnormalized {\em LZ
complexity} of $x^k$ whereas the normalized LZ complexity is defined as 
\begin{equation}
\rho_{\mbox{\tiny LZ}}(x^k)\dfn
\frac{c(x^k)\log
c(x^k)}{k}.
\end{equation}

A useful inequality, that relates the empirical entropy of non-overlapping
$\ell$-blocks of $x^k$ (where $\ell$ divides $k$) and $\rho_{\mbox{\tiny LZ}}(x^k)$ (see, for example, 
\cite[eq.\ (26)]{me23}), is the following:
\begin{eqnarray}
\label{zivineq}
\frac{\hH(X^\ell)}{\ell}&\ge&\rho_{\mbox{\tiny LZ}}(x^k)
-\frac{\log[4S^2(\ell)]\log\alpha}{(1-\varepsilon_k)\log
k}-\frac{S^2(\ell)\log[4S^2(\ell)]}{k}-\frac{1}{\ell}\nonumber\\
&\dfn&\rho_{\mbox{\tiny LZ}}(x^k)-\Delta_k(\ell),
\end{eqnarray}
where
\begin{equation}
S(\ell)=\sum_{i=0}^{\ell-1}\alpha^i=\frac{\alpha^\ell-1}{\alpha-1}.
\end{equation}
It is obtained from the fact that the Shannon code for $\ell$-blocks can be
implemented using a finite-state encoder with no more than $S(\ell)$
states\footnote{For a block code of length $\ell$ to be implemented by a
finite-state machine, one defines the state at each time instant $i$ to be the
contents of the input, starting at the beginning of the current block (at time
$\ell\cdot\lfloor i/\ell\rfloor+1$) and ending at time $i-1$. The number of states
for an input alphabet of size $\alpha$ is then
$\sum_{i=0}^{\ell-1}\alpha^i=(\alpha^\ell-1)/(\alpha-1)<\alpha^\ell$.}
and therefore it must comply with the lower bound of \cite[Theorem 1]{ZL78}.
Note that $\lim_{k\to\infty}\Delta_k(\ell)=1/\ell$ and so,
$\lim_{\ell\to\infty}\lim_{k\to\infty}\Delta_k(\ell)=0$. Clearly, it is
possible to let $\ell=\ell(k)$ increase with $k$ slowly enough such that
$\Delta_k(\ell(k))\to 0$ as $k\to\infty$, in particular, $\ell(k)$ should be $o(\log k)$
for that purpose.

In \cite{Ziv85}, the notion of the LZ complexity was extended to incorporate
finite-state lossless compression in the presence of side information, namely,
the conditional version of the LZ complexity.
Given $x^k$ and $y^k$,
let us apply the incremental
parsing procedure of the LZ algorithm
to the sequence of pairs $((x_1,y_1),(x_2,y_2),\ldots,(x_k,y_k))$.
As mentioned before, according to this procedure, all phrases are distinct
with a possible exception of the last phrase, which might be incomplete.
Let $c(x^k,y^k)$ denote the number of distinct phrases.
For example,\footnote{The same example appears in \cite{Ziv85}.} if
\begin{eqnarray}
x^6&=&0~|~1~|~0~0~|~0~1|\nonumber\\
y^6&=&0~|~1~|~0~1~|~0~1|\nonumber
\end{eqnarray}
then $c(x^6,y^6)=4$.
Let $c(y^k)$ denote the resulting number of distinct phrases
of $y^k$, and let $y(l)$ denote the $l$-th distinct $y$--phrase,
$l=1,2,\ldots,c(y^k)$. In the above example, $c(y^6)=3$. Denote by
$c_l(x^k|y^k)$ the number of occurrences of $y(l)$ in the
parsing of $y^k$, or equivalently, the number of distinct $x$-phrases
that jointly appear with $y(l)$. Clearly, $\sum_{l=1}^{c(y^k)} c_l(x^k|y^k)=
c(x^k,y^k)$. In the above example, $y(1)=0$, $y(2)=1$, $y(3)=01$,
$c_1(x^6|y^6)=c_2(x^6|y^6)=1$, and $c_3(x^6|y^6)=2$. Now, the conditional LZ
complexity of $x^k$ given $y^k$ is defined as
\begin{equation}
\rho_{LZ}(x^k|y^k)\dfn\frac{1}{k}\sum_{l=1}^{c(y^k)}c_l(x^k|y^k)\log
c_l(x^k|y^k).
\end{equation}
In \cite{Ziv85} it was shown that $\rho_{LZ}(x^k|y^k)$ is the main term of the
compression ratio achieved by the conditional version of the LZ algorithm
described therein (see
also \cite{UK03}), i.e., the length function, $LZ(x^k|y^k)$, of the coding scheme
proposed therein is upper bounded (in parallel to (\ref{lz-clogc})) by 
\begin{equation}
\label{conditional-lz}
LZ(x^k|y^k)\le k\rho_{LZ}(x^k|y^k)+k\hat{\epsilon}(k),
\end{equation}
where $\hat{\epsilon}(k)$ is a certain sequence that tends to zero as $k\to\infty$.
On the other hand, analogously to \cite[Theorem 1]{ZL78}, it was shown in
\cite{me00}, that $\rho_{LZ}(x^k|y^k)$ is also the main term of a lower bound
to the compression ratio that can be achieved by any finite-state encoder with
side information at both ends, provided that the number of states is not
too large, similarly as described above for the unconditional version. 

The inequality (\ref{zivineq}) also extends to the conditional case as
follows (see \cite{me00}):
\begin{equation}
\label{conditionalzivineq}
\frac{\hH(X^\ell|Y^\ell)}{\ell}
\ge\rho_{\mbox{\tiny LZ}}(x^k|y^k)-\Delta_k'(\ell),
\end{equation}
where $\Delta_k'(\ell)$ is the same as $\Delta_k(\ell)$ except that
the expression of $S(\ell)$ included therein is redefined as
$(\alpha^\ell\beta^\ell-1)/(\alpha\beta-1)$ to accommodate the number of
states associated with the conditional version of the aforementioned Shannon code applied to
$\ell$-blocks. By the same token, we also have
\begin{equation}
\frac{\hH(X^\ell,Y^\ell)}{\ell}
\ge\rho_{\mbox{\tiny LZ}}(x^k,y^k)-\Delta_k'(\ell).
\end{equation}

We close this section with a comment that although $\rho_{\mbox{\tiny
LZ}}(y^k)$ and $\rho_{\mbox{\tiny LZ}}(x^k|y^k)$ can be thought of as
deterministic counterparts of the entropies and conditional entropies
\cite{Ziv85}, \cite{ZL78}, 
to the best knowledge of the author,
there is no apparent parallel ``chain rule'' that explicitly decomposes
$\rho_{\mbox{\tiny LZ}}(x^k,y^k)$ as $\rho_{\mbox{\tiny
LZ}}(y^k)+\rho_{\mbox{\tiny LZ}}(x^k|y^k)$ or as
$\rho_{\mbox{\tiny
LZ}}(x^k)+\rho_{\mbox{\tiny LZ}}(y^k|x^k)$. However, we will be able to
establish a certain relationship in 
this spirit at least some asymptotic sense.\footnote{
It is interesting to note, in this context, that the Kolmogorov complexity
also obeys a parallel chain rule in a certain asymptotic sense, as asserted by the
Kolmogorov-Levin theorem \cite{Kolmogorov68}, \cite{ZL70}.}
As described in the Introduction, this will be instrumental in establishing
the ``corner points'' of the achievable rate region.

\section{The Coding Theorem}
\label{codingtheorem}

In \cite{ZL78} there is a coding theorem and a matching converse in terms of
the long-term average of $\rho{\mbox{\tiny LZ}}(\cdot)$ applied in successive
non-overlapping blocks of the infinite sequence $\bx$, namely,
\begin{equation}
\rho(\bx)=\limsup_{k\to\infty}\limsup_{n\to\infty}\frac{k}{n}\sum_{i=0}^{n/k-1}\rho_{\mbox{\tiny
LZ}}(x_{ik+1}^{ik+k}),
\end{equation}
which is a worst-case approach, where the compression ratio is probed at a
sequence of block lengths with the
worst possible limit.

A similar approach will be executed here too.
In particular, denoting
\begin{eqnarray}
\rho_k(x^n,y^n)&=&\frac{k}{n}\sum_{i=0}^{n/k-1}\rho_{\mbox{\tiny
LZ}}(x_{ik+1}^{ik+k},y_{ik+1}^{ik+k}),\\
\rho_k(x^n|y^n)&=&\frac{k}{n}\sum_{i=0}^{n/k-1}\rho_{\mbox{\tiny
LZ}}(x_{ik+1}^{ik+k}|y_{ik+1}^{ik+k}),\\
\rho_k(y^n|x^n)&=&\frac{k}{n}\sum_{i=0}^{n/k-1}\rho_{\mbox{\tiny
LZ}}(y_{ik+1}^{ik+k}|x_{ik+1}^{ik+k}),
\end{eqnarray}
we define the quantities
\begin{eqnarray}
\rho(\bx,\by)&=&\limsup_{k\to\infty}\limsup_{n\to\infty}\rho_k(x^n,y^n)\\
\rho(\bx|\by)&=&\limsup_{k\to\infty}\limsup_{n\to\infty}\rho_k(x^n|y^n)\\
\rho(\by|\bx)&=&\limsup_{k\to\infty}\limsup_{n\to\infty}\rho_k(y^n|x^n).
\end{eqnarray}

\begin{theorem}
\label{thm1}
Consider the setting defined in Subsection \ref{formulation}
and let $\calR(\bx,\by)$ be defined as in Subsection \ref{objectives}.
Then,
\begin{equation}
\calR(\bx,\by)=R(\bx,\by)\dfn
\{(R_{\mbox{\tiny x}},R_{\mbox{\tiny y}}):~R_{\mbox{\tiny x}}\ge
\rho(\bx|\by),~R_{\mbox{\tiny y}}\ge\rho(\by|\bx),~R_{\mbox{\tiny x}}+
R_{\mbox{\tiny y}}\ge\rho(\bx,\by)\}.
\end{equation}
\end{theorem}

The converse part of Theorem \ref{thm1}, asserting that
$\calR(\bx,\by)\subseteq R(\bx,\by)$, is proved in Appendix A, and
the direct part, asserting that
$R(\bx,\by)\subseteq \calR(\bx,\by)$, is proved in Appendix B.\\

A natural question that arises with respect to the coding theorem concerns
the corner points of the rate region. If $R_{\mbox{\tiny
x}}=\rho(\bx)$ and $R_{\mbox{\tiny x}}+R_{\mbox{\tiny y}}=\rho(\bx,\by)$,
which is one of the corner points at the boundary of the achievable region,
then $R_{\mbox{\tiny y}}=\rho(\bx,\by)-\rho(\bx)$. In analogy to
the traditional probabilistic setting, where $H(X,Y)-H(X)=H(Y|X)$, it is natural to expect that
$R_{\mbox{\tiny y}}=\rho(\by|\bx)$. This expectation could be met if we can
establish a ``chain rule'',
$\rho(\bx,\by)=\rho(\bx)+\rho(\by|\bx)$.
While there is no known chain rule for the LZ complexities of finite source
strings, it turns out that in the asymptotic limit, such a chain rule actually
applies in a certain sense. To this end, we assert the following theorem,
whose proof appears in Appendix C.

\begin{theorem}
\label{thm2}
Define
\begin{eqnarray}
\rho_{\mbox{\tiny LZ}}^+(x^k,y^k)&=&\max\{\rho_{\mbox{\tiny
LZ}}(x^k,y^k),\rho_{\mbox{\tiny LZ}}(x^k)+\rho_{\mbox{\tiny
LZ}}(y^k|x^k),\rho_{\mbox{\tiny LZ}}(y^k)+\rho_{\mbox{\tiny LZ}}(x^k|y^k)\},\\
\rho_{\mbox{\tiny LZ}}^-(x^k,y^k)&=&\min\{\rho_{\mbox{\tiny
LZ}}(x^k,y^k),\rho_{\mbox{\tiny LZ}}(x^k)+\rho_{\mbox{\tiny
LZ}}(y^k|x^k),\rho_{\mbox{\tiny LZ}}(y^k)+\rho_{\mbox{\tiny LZ}}(x^k|y^k)\}.
\end{eqnarray}
Given $\bx$ and $\by$, let
\begin{eqnarray}
\rho^+(\bx,\by)&=&\limsup_{k\to\infty}\limsup_{n\to\infty}\frac{k}{n}\sum_{i=0}^{n/k-1}\rho_{\mbox{\tiny
LZ}}^+(x_{ik+1}^{ik+k},y_{ik+1}^{ik+k})\\
\rho^-(\bx,\by)&=&\limsup_{k\to\infty}\limsup_{n\to\infty}\frac{k}{n}\sum_{i=0}^{n/k-1}\rho_{\mbox{\tiny
LZ}}^-(x_{ik+1}^{ik+k},y_{ik+1}^{ik+k}).
\end{eqnarray}
Then, 
\begin{equation}
\label{equalrhos}
\rho^+(\bx,\by)=\rho^-(\bx,\by)=\rho(\bx,\by).
\end{equation}
\end{theorem}

We remark in passing that, as can be seen in the proof of Theorem \ref{thm2}, 
the three equivalent quantities of eq.\
(\ref{equalrhos}) are also equal to yet another important well-known quantity,
which is the finite-state compressibility of $(x^n,y^n)$, denoted here
by $\varrho_\infty(\bx,\by)$ \cite[eq.\ (4)]{ZL78}.
For convenience, we
remind here the definition of the finite-state compressibility in a few
steps: Let $\varrho_s(x^n,y^n)$ denote the minimum compression ratio achieved by
any information lossless $s$-state (joint) encoder on $(x^n,y^n)$. Next,
define $\varrho_s(\bx,\by)=\limsup_{n\to\infty}\varrho_s(x^n,y^n)$, and
finally, $\varrho_\infty(\bx,\by)=\lim_{s\to\infty}\varrho_s(\bx,\by)$.

To see why the chain rule,
$\rho(\bx,\by)=\rho(\bx)+\rho(\by|\bx)$,
holds true, consider the following argument. First, observe that
\begin{eqnarray}
\rho(\bx,\by)&=&\rho^-(\bx,\by)\nonumber\\
&\le&\limsup_{k\to\infty}\limsup_{n\to\infty}\frac{k}{n}\sum_{i=0}^{n/k-1}[\rho_{\mbox{\tiny
LZ}}(x^k)+\rho_{\mbox{\tiny
LZ}}(y^k|x^k)]\nonumber\\
&\le&\limsup_{k\to\infty}\limsup_{n\to\infty}\frac{k}{n}\sum_{i=0}^{n/k-1}\rho_{\mbox{\tiny
LZ}}(x^k)+\limsup_{k\to\infty}\limsup_{n\to\infty}\frac{k}{n}\sum_{i=0}^{n/k-1}\rho_{\mbox{\tiny
LZ}}(y^k|x^k)\nonumber\\
&=&\rho(\bx)+\rho(\by|\bx).
\end{eqnarray}
To establish the reverse inequality,
$\rho(\bx,\by)\ge\rho(\bx)+\rho(\by|\bx)$ (and hence equality), consider the following:
On the one hand, given that $R_{\mbox{\tiny x}}=\rho(\bx)$, then by the
direct part of Theorem \ref{thm1}, the rate
$R_{\mbox{\tiny y}}=\rho(\bx,\by)-\rho(\bx)$ is achievable for the y-encoder,
as the point $(\rho(\bx),\rho(\bx,\by)-\rho(\bx))$ is at (the boundary of) the
achievable rate region. On the other hand, given that $R_{\mbox{\tiny
x}}=\rho(\bx)$, the source sequence $\bx$ becomes side information that is available at both ends, and
then the compression ratio for $\by$ in the presence of $\bx$ is lower bounded by the conditional
LZ complexity, $\rho(\by|\bx)$ \cite{me00}. Therefore,
$\rho(\bx,\by)-\rho(\bx)\ge\rho(\by|\bx)$.

\noindent
{\bf Discussion.}\\
Several aspects of Theorem \ref{thm1} should be highlighted.\\

\noindent
1. It is natural to compare our results to those of Ziv in \cite{Ziv84}.
As mentioned in the Introduction,
the class of communication systems allowed here is
somewhat broader. We allow
variable-rate codes, as opposed to fixed-rate codes
of \cite{Ziv84}. Also, in our setting there are no limitations on the decoder
whereas in \cite{Ziv84}, a finite-state decoder is assumed. Finally, we
consider the complete setting of SW coding, where both sources are compressed,
whereas in \cite{Ziv84}, only one source is compressed and the other source
serves as side information. Finally, our coding theorem is more closely
related to those of \cite{Ziv85} and \cite{ZL78}, since it is about a similar
form of the conditional LZ complexity. We show that the conditional LZ
complexity, $\rho_{\mbox{\tiny LZ}}(\bx|\by)$, is operatively meaningful, not
only when the side information is available at both ends, but also when it is
available at the decoder only. Similarly as in \cite{Ziv84}, our direct
theorem asserts that the expected Hamming distortion is asymptotically vanishing,
where the expectation is w.r.t.\ the randomness of the code in the ensemble. 
Since the code is re-selected in every block independently,
for an infinite sequence pair, $(\bx,\by)$, the Hamming distortion eventually
vanishes almost surely.\\

\noindent
2. As can be seen in the proof of the direct part of Theorem \ref{thm1}, we
apply in each block of length $k$ a universal decoder that maximizes the
decoding metric,
\begin{equation}
u(x^k,y^k)=\min\{[R_{\mbox{\tiny x}}-\rho_{\mbox{\tiny LZ}}(x^k|y^k)], 
[R_{\mbox{\tiny y}}-\rho_{\mbox{\tiny LZ}}(y^k|x^k)], 
[R_{\mbox{\tiny x}}+R_{\mbox{\tiny y}}-\rho_{\mbox{\tiny LZ}}(x^k,y^k)]\}
\end{equation}
among all pairs of vectors that are consistent with the given bin assignments.
Note that it is composed of three
universal decoding metrics, each one of which handles a
different type of error event: (i) error in $x^k$ only, (ii) error in $y^k$ only, and
(iii) error in both $x^k$ and $y^k$. Note that this is different from the universal
decoder of Csisz\'ar and K\"orner for memoryless sources \cite[Problem 13.6(b), page 267]{CK11},
which simply minimizes the joint empirical entropy. The reason that the
empirical joint entropy handles successfully all three types of errors is
associated with the fact that the empirical entropy satisfies the chain rule,
$\hat{H}(X,Y)=\hat{H}(X)+\hat{H}(Y|X)=\hat{H}(Y)+\hat{H}(X|Y)$, and so, for errors
of types (i) and (ii) the minimum joint entropy decoder is equivalent to minimizing
$\hat{H}(X|Y)$ or 
$\hat{H}(Y|X)$, respectively. Here, on the other hand, this is not the case, because
as mentioned before, there is no apparent chain rule for LZ compression ratios
for vectors of finite length.
Therefore, three different metrics are required.\\

\noindent
3. Another interesting observation about our universal decoder is the
following. As is well known, in SW decoding each bin functions like a channel code. Since we are
talking about universal decoding, it is not surprising to see here universal
decoding metrics in the spirit of the maximum mutual information (MMI) decoder \cite{CK11} or the minimum conditional
entropy decoder, or Ziv's 1985 universal decoding metric \cite{Ziv85}. What
seems to be less trivial is the fact that these universal decoding metrics
continue to work well even in the present context
of individual sequences, as in contrast to the setting of \cite{Ziv85}, here there
is no finite-state channel that relates $\by$ to $\bx$, and there is no random
coding behind the channel inputs.\\

\noindent
4. In the direct part, we use a fixed-rate SW code within each $k$-block.
Clearly, it is problematic
to use a fixed-rate code since the joint `statistics' of $\bx$ and $\by$ are
not known ahead of time and it is not possible to know in advance the joint LZ
complexities and the conditional LZ complexities within each such block in
order to assign coding rates accordingly.
However, with very little feedback from the decoder
to the encoder, one could construct an adaptive mechanism that in some way `learns' the
joint statistics. One such scheme, which is a modified version of Draper's
scheme \cite{Draper04} is proposed in the next section.

\section{Incremental SW coding}
\label{incremental}

In \cite{Draper04}, Draper proposed and analyzed a universal incremental SW
coding scheme for memoryless sources. It turns out that this scheme can be
used almost verbatim in the individual-sequence setting considered here, provided
that some adjustments are made. Note that our notation here is
somewhat different from that of \cite{Draper04}.

Draper assumed that the x-encoder (resp.\ y-encoder) are both connected 
to fixed-rate noiseless channels (bit pipes), and that
the x-encoder (resp.\ y-encoder) communicates $r_{\mbox{\tiny
x}}$ (resp.\ $r_{\mbox{\tiny y}}$) bits per channel use. More precisely, after $m$
channel uses, the x-encoder (resp.\ y-encoder) has transmitted 
$\lfloor mr_{\mbox{\tiny x}}\rfloor$ (resp.\
$\lfloor mr_{\mbox{\tiny
y}}\rfloor$) bits over the channel. 
The proposed coding scheme works in the following
stages:

\begin{enumerate}
\item The x-encoder (resp.\ y-encoder) observes the full block, $x^k$ (resp.\
$y^k$) and calculates $\rho_{\mbox{\tiny LZ}}(x^k)$ (resp.\ $\rho_{\mbox{\tiny
LZ}}(y^k)$).
\item The x-encoder (resp.\ y-encoder) communicates the value of
$\rho_{\mbox{\tiny LZ}}(x^k)$ (resp.\ $\rho_{\mbox{\tiny
LZ}}(y^k)$) as a header, using approximately $\log n$ bits.
Let $\calT(x^k)=\{\tx^k:~\rho_{\mbox{\tiny LZ}}(\tx^k)=\rho_{\mbox{\tiny
LZ}}(x^k)\}$ and 
$\calT(y^k)=\{\ty^k:~\rho_{\mbox{\tiny LZ}}(\ty^k)=\rho_{\mbox{\tiny
LZ}}(y^k)\}$ denote the ``type classes'' of $x^k$ and $y^k$, respectively. 
For each possible type class, the respective encoder and decoder agree (ahead of time) on
the order of the list of members of that type class. This order is selected
independently at random for each type class of each source.
\item The x-encoder (resp.\ y-encoder) sequentially transmits successive bits
of the binary expansion of the location of $x^k$ (resp.\ $y^k$) in the shared
list. After $m$ channel uses, the decoder has received the first
$\lfloor mr_{\mbox{\tiny x}}\rfloor$ (resp.\ 
$\lfloor mr_{\mbox{\tiny y}}\rfloor$) of the location of $x^k$ (resp.\ $y^k$)
in the list. Each incomplete binary expansion corresponds to a bin of
sequences whose locations in the list share the same most significant bits. 
Since the binary expansions are nested, the bins are nested as well.
Referring to the x-encoder (and similarly, for the y-encoder),
initially, the bin $\calB_0(x^k)$ is the entire type class, $\calT(x^k)$. After the first channel use,
the bin shrinks to become $\calB_1(x^k)$, which is the set of sequences whose position index share the same 
first $\lfloor r_{\mbox{\tiny x}}\rfloor$ bits. After the second channel use,
it shrinks further to $\calB_2(x^k)$, which is the set with the same $\lfloor 2r_{\mbox{\tiny
x}}\rfloor$ most significant bits, and so on.
\item After each channel use, the decoder tests all pairs of sequences
$(\tx^k,\ty^k)$ that are consistent with the corresponding bins currently known to the
decoder. Specifically, after $m$ channel uses, it compares an empirical mutual information,
$\hat{I}_m(x^k;y^k)$ (to be defined shortly) to a time-varying threshold,
$\theta_m$ (to be defined shortly as well). As soon as
$\hat{I}_m(\tx^k;\ty^k)\ge\theta_m$ for some vector pair, the decoder sends an {\tt ACK} to the encoders,
which then cease to transmit.
\item If the x-encoder (resp.\ y-encoder) has already transmitted
$k\rho_{\mbox{\tiny LZ}}(x^k)$ bits (resp.\
$k\rho_{\mbox{\tiny LZ}}(y^k)$ bits), excluding the header,
it ceases to transmit even it has not
yet received an {\tt ACK}. 
\end{enumerate}
Now, for $m=1,2,\ldots$, define
\begin{equation}
\hat{I}_m(x^k;y^k)=\left\{\begin{array}{ll}
\rho_{\mbox{\tiny LZ}}(x^k)+\rho_{\mbox{\tiny LZ}}(y^k)-\rho_{\mbox{\tiny
LZ}}(x^k,y^k) & mr_{\mbox{\tiny x}}<LZ(x^k),~mr_{\mbox{\tiny y}}<LZ(y^k)\\
\rho_{\mbox{\tiny LZ}}(x^k)-\rho_{\mbox{\tiny
LZ}}(x^k|y^k) & mr_{\mbox{\tiny x}}<LZ(x^k),~mr_{\mbox{\tiny y}}\ge LZ(y^k)\\
\rho_{\mbox{\tiny LZ}}(y^k)-\rho_{\mbox{\tiny
LZ}}(y^k|x^k) & mr_{\mbox{\tiny x}}\ge LZ(x^k),~mr_{\mbox{\tiny
y}}<LZ(y^k)\end{array}\right.
\end{equation}
and
\begin{equation}
\theta_m=\frac{[LZ(x^k)-mr_{\mbox{\tiny x}}]_++[LZ(y^k)-mr_{\mbox{\tiny
y}})]_+}{k}+\epsilon,
\end{equation}
where $\epsilon>0$ is arbitrarily small.

To analyze the performance of this coding scheme, we proceed similarly as in
\cite{Draper04}, but with a few twists.
We first define the error event after $m$ channel uses:
\begin{equation}
\calE_m=\{(\tx^k,\ty^k)\ne(x^k,y^k):~\rho_{\mbox{\tiny
LZ}}(\tx^k)=\rho_{\mbox{\tiny LZ}}(x^k),~\rho_{\mbox{\tiny
LZ}}(\ty^k)=\rho_{\mbox{\tiny LZ}}(y^k),~\hat{I}_m(\tx^k;\ty^k)\ge\theta_m\},
\end{equation}
and three critical values of $m$:
\begin{eqnarray}
m_{\mbox{\tiny x}}&=&\frac{LZ(x^k)}{r_{\mbox{\tiny x}}},\\
m_{\mbox{\tiny y}}&=&\frac{LZ(y^k)}{r_{\mbox{\tiny y}}},\\
m_{\mbox{\tiny xy}}&=&\frac{LZ(x^k,y^k)}{r_{\mbox{\tiny x}}+r_{\mbox{\tiny y}}}.
\end{eqnarray}
Now, there are three cases, according to the smallest number among
$m_{\mbox{\tiny x}}$,
$m_{\mbox{\tiny y}}$, and
$m_{\mbox{\tiny xy}}$: (i)
$m_{\mbox{\tiny xy}}<\min\{m_{\mbox{\tiny x}},m_{\mbox{\tiny y}}\}$, (ii)
$m_{\mbox{\tiny x}}<\min\{m_{\mbox{\tiny y}},m_{\mbox{\tiny xy}}\}$, and (iii)
$m_{\mbox{\tiny y}}<\min\{m_{\mbox{\tiny x}},m_{\mbox{\tiny xy}}\}$.

Consider case (i) first.
As long as $m<m_{\mbox{\tiny xy}}$, 
the probability of error after $m$
channel uses, is upper bounded by
\begin{eqnarray}
\mbox{Pr}\{\calE_m\}&=&\sum_{(\tx^k,\ty^k)\in\calE_m}2^{-\lfloor mr_{\mbox{\tiny
x}}\rfloor-\lfloor mr_{\mbox{\tiny y}}\rfloor}\nonumber\\
&\le&4\cdot\sum_{\{(\tx^k,\ty^k):~LZ(\tx^k,\ty^k)\le LZ(x^k)+LZ(y^k)-k\theta_m\}}2^{-m(r_{\mbox{\tiny
x}}+r_{\mbox{\tiny y}})}\nonumber\\
&\le&4\cdot\sum_{\{(\tx^k,\ty^k):~LZ(\tx^k,\ty^k)\le LZ(x^k)+LZ(y^k)-k\theta_m\}}2^{-m(r_{\mbox{\tiny
x}}+r_{\mbox{\tiny y}})}\nonumber\\
&\le&8\cdot 2^{LZ(x^k)+LZ(y^k)-k\theta_m}\cdot 2^{-m(r_{\mbox{\tiny
x}}+r_{\mbox{\tiny y}})}\nonumber\\
&=& 8\cdot 2^{-k\epsilon},
\end{eqnarray}
where the last step follows from the same argument as in
step (c) of eq.\ (\ref{B10}) (see (\ref{B11})).
After $m$ exceeds $m_{\mbox{\tiny xy}}$, at least the correct pair,
$(x^k,y^k)$,
certainly exceeds
the threshold, but by that time, the two encoders together have transmitted
just above $LZ(x^k,y^k)$ bits. 

In case (ii), as long as $m< m_{\mbox{\tiny x}}$, the probability of error is the same as
before. Once $m$ exceeds $m_{\mbox{\tiny x}}$ (and both encoder and decoder know when this
happens as they both know $LZ(x^k)$ and $r_{\mbox{\tiny x}}$), the
x-encoder may cease to transmit and the decoder can decode $x^k$ with high
reliability by seeking a vector $\tx^k$ such that:
(i) $LZ(\tx^k)$ agrees
with the given $LZ(x^k)$, and 
(ii) the first $mr_{\mbox{\tiny x}}$ bits of the bin index of $\tx^k$ agree with those that have
been received at the decoder from the x-encoder. With high probability there
is only one such $\tx^k$ and it then
must by the correct $x^k$. At this point, only the transmission of the
y-encoder continues. Assuming the $x^k$ was decoded correctly, the
probability of error after $m$ steps is now upper bounded by:
\begin{eqnarray}
\mbox{Pr}\{\calE_m\}&=&\sum_{\{\ty^k:~(x^k,\ty^k)\in\calE_m}2^{-\lfloor mr_{\mbox{\tiny
y}}\rfloor}\nonumber\\
&\le&2\cdot\sum_{\{\ty^k:~LZ(\ty^k|x^k)\le
LZ(y^k)-k\theta_m\}}2^{-m
r_{\mbox{\tiny y}}}\nonumber\\
&\le&4\cdot 2^{LZ(y^k)-k\theta_m}\cdot 2^{-mr_{\mbox{\tiny
y}}}\nonumber\\
&=& 4\cdot 2^{-k\epsilon}.
\end{eqnarray}
As soon as $mr_{\mbox{\tiny y}}$ exceeds $LZ(y^k|x^k)$, the correct pair,
$(x^k,y^k)$ exceeds the threshold. At this time, the transmission of the y-encoder
stops too and the decoding of $y^k$ can be carried out with $x^k$ in the role
of decoder side information.
Case (iii) is exactly like case (ii), except that the roles of $x^k$ and $y^k$ are
swapped. 

\section*{Appendix A}
\renewcommand{\theequation}{A.\arabic{equation}}
    \setcounter{equation}{0}

\noindent
{\em Proof of the converse part of Theorem \ref{thm1}.}\\
We first extend the generalized Kraft inequality of \cite[Lemma 2]{ZL78} from
information lossless encoders to $\epsilon$-lossy encoders.
Specifically, we argue that for any given $\epsilon$-lossy encoder pair with
$q^2$ states (i.e., $q$ states of $E_{\mbox{\tiny x}}$ times $q$ states of $E_{\mbox{\tiny
y}}$),
\begin{equation}
\label{kraft}
\sum_{(x^\ell,y^\ell)\in\calX^\ell\times\calY^\ell}2^{-\{\min_{s\in\calS}L[f_{\mbox{\tiny
x}}(s,x^\ell,y^\ell)]+\min_{z\in\calZ}L[f_{\mbox{\tiny y}}(z,y^\ell,x^\ell)]\}}\le
q^4B_\ell(\epsilon)\left(1+\log\left[1+\frac{\alpha^\ell\beta^\ell}{q^4B_\ell(\epsilon)}\right]\right),
\end{equation}
where
\begin{equation}
B_\ell(\epsilon)=\sum_{j=0}^{\ell\epsilon}\left(\begin{array}{cc}
l\\j\end{array}\right)(\alpha\beta-1)^j
\end{equation}
is the size of the Hamming sphere of radius $\ell\epsilon$ in the space
$\calX^\ell\times\calY^\ell$ whose size is $\alpha^\ell\beta^\ell$. Using the
Chernoff bound, it can be readily seen that
\begin{equation}
B_\ell(\epsilon)\le 2^{\ell Q(\epsilon)}
\end{equation}
where
\begin{equation}
Q(\epsilon)=\left\{\begin{array}{ll}
h_2(\epsilon)+\epsilon\log(\alpha\beta-1) & \epsilon <
1-\frac{1}{\alpha\beta}\\
\log(\alpha\beta) & 1-\frac{1}{\alpha\beta}\le\epsilon\le 1\end{array}\right.
\end{equation}
and where $h_2(\epsilon)=-\epsilon\log\epsilon-(1-\epsilon)\log(1-\epsilon)$ is
the binary entropy function. Since we consider small values of $\epsilon$ the
second line in the definition of $Q(\epsilon)$ will not be relevant to our
derivations. We henceforth denote
\begin{equation}
\delta(\epsilon)\dfn h_2(\epsilon)+\epsilon\log(\alpha\beta-1).
\end{equation}
The proof of eq.\ (\ref{kraft}) is exactly the same as the proof of
\cite[Lemma 2]{ZL78},
where the only modification needed is that here, the number $k_j$ of $(x^\ell,y^\ell)$ for which
$\min_{s\in\calS}L[f_{\mbox{\tiny x}}(s,x^\ell,y^\ell)]+\min_{z\in\calZ}L[f_{\mbox{\tiny
y}}(z,y^\ell,x^\ell)]=j$ is upper bounded by $q^4B_\ell(\epsilon)2^j$, as follows
from the postulate that the encoders form an $\epsilon$-lossy system.
It follows that the total description length at the outputs of the encoders is
lower bounded as follows.
\begin{eqnarray}
n(R_{\mbox{\tiny x}}+R_{\mbox{\tiny y}})&\ge&\sum_{t=1}^n \{L[f_{\mbox{\tiny
x}}(s_t,x_t,y_t)]+L[f_{\mbox{\tiny y}}(z_t,y_t,x_t)]\}\nonumber\\
&=&\sum_{i=0}^{n/k-1}\sum_{m=0}^{k/\ell-1} \sum_{j=1}^\ell
\{L[f_{\mbox{\tiny x}}(s_i,x_{ik+m\ell+j},y_{ik+m\ell+j}]+
L[f_{\mbox{\tiny y}}(z_i,y_{ik+m\ell+j},x_{ik+m\ell+j}]\}\nonumber\\
&=&\sum_{i=0}^{n/k-1}\sum_{m=0}^{k/\ell-1}
\{L[f_{\mbox{\tiny
x}}(s_i,x_{ik+m\ell+1}^{ik+m\ell+\ell},y_{ik+m\ell+1}^{ik+m\ell+\ell})]+
L[f_{\mbox{\tiny
y}}(z_i,y_{ik+m\ell+1}^{ik+m\ell+\ell},x_{ik+m\ell+1}^{ik+m\ell+\ell})]\}\nonumber\\
&\ge&\sum_{i=0}^{n/k-1}\sum_{m=0}^{k/\ell-1}
\{\min_{s\in\calS}L[f_{\mbox{\tiny
x}}(s,x_{ik+m\ell+1}^{ik+m\ell+\ell},y_{ik+m\ell+1}^{ik+m\ell+\ell})]+
\min_{z\in\calZ}L[f_{\mbox{\tiny
y}}(z,y_{ik+m\ell+1}^{ik+m\ell+\ell},x_{ik+m\ell+1}^{ik+m\ell+\ell})]\}\nonumber\\
&=&\sum_{i=0}^{n/k-1}\frac{k}{\ell}
\sum_{(x^\ell,y^\ell)\in\calX^\ell\times\calY^\ell}\hat{P}_i(x^\ell,y^\ell)\cdot[\min_{s\in\calS}L[f_{\mbox{\tiny
x}}(s,x^\ell,y^\ell)]+\min_{z\in\calZ}L[f_{\mbox{\tiny y}}(z,y^\ell,x^\ell)].
\end{eqnarray}
Now, according to the generalized Kraft inequality,
\begin{eqnarray}
& &q^4B_\ell(\epsilon)\left(1+\log\left[1+\frac{(\alpha\beta)^\ell}
{q^4B_\ell(\epsilon)}\right]\right)\nonumber\\&\ge&
\sum_{(x^\ell,y^\ell)\in\calX^\ell\times\calY^\ell}
\exp_2\{-[\min_{s\in\calS}L[f_{\mbox{\tiny
x}}(s,x^\ell,y^\ell)]+\min_{z\in\calZ}L[f_{\mbox{\tiny y}}(z,y^\ell,x^\ell)]\}\nonumber\\
&=&\sum_{(x^\ell,y^\ell)\in\calX^\ell\times\calY^\ell}
\hat{P}_i(x^\ell,y^\ell)\cdot
\exp_2\{-[\min_{s\in\calS}L[f_{\mbox{\tiny
x}}(s,x^\ell,y^\ell)]+\min_{z\in\calZ}L[f_{\mbox{\tiny y}}(z,y^\ell,x^\ell)]-\log
\hat{P}_i(x^\ell,y^\ell)\}\nonumber\\
&\ge&\exp_2\bigg\{-\sum_{(x^\ell,y^\ell)\in\calX^\ell\times\calY^\ell}\hat{P}_i(x^\ell,y^\ell)\cdot
\bigg(\min_{s\in\calS}L[f_{\mbox{\tiny
x}}(s,x^\ell,y^\ell)]+\nonumber\\
& &\min_{z\in\calZ}L[f_{\mbox{\tiny y}}(z,y^\ell,x^\ell)]\bigg)
+\hH(X_i^\ell,Y_i^\ell)\bigg\}.
\end{eqnarray}
Taking the base 2 logarithms of both sides, this yields
\begin{eqnarray}
& &\log\left\{q^4B_\ell(\epsilon)\left(1+\log\left[1+\frac{(\alpha\beta)^\ell}{q^4B_\ell(\epsilon)}\right]\right)\right\}\nonumber\\
&\ge&\hH(X_i^\ell,Y_i^\ell)-\sum_{(x^\ell,y^\ell)\in\calX^\ell\times\calY^\ell}\hat{P}_i(x^\ell,y^\ell)\cdot
[\min_{s\in\calS}L[f_{\mbox{\tiny
x}}(s,x^\ell,y^\ell)]+\min_{z\in\calZ}L[f_{\mbox{\tiny y}}(z,y^\ell,x^\ell)],
\end{eqnarray}
implying that
\begin{eqnarray}
\label{Rx+Ry}
& &R_{\mbox{\tiny x}}+R_{\mbox{\tiny
y}}\nonumber\\
&\ge&\frac{k}{n}\sum_{i=0}^{n/k-1}\frac{1}{\ell}\sum_{(x^\ell,y^\ell)\in\calX^\ell\times\calY^\ell}
\hat{P}_i(x^\ell,y^\ell)\cdot
\bigg(\min_{s\in\calS}L[f_{\mbox{\tiny
x}}(s,x^\ell,y^\ell)]+\min_{z\in\calZ}L[f_{\mbox{\tiny
y}}(z,y^\ell,x^\ell)]\bigg)\nonumber\\
&\ge&\frac{k}{n}\sum_{i=0}^{n/k-1}\frac{\hat{H}(X_i^\ell,Y_i^\ell)}{\ell}-
\frac{1}{\ell}\log\left\{q^4B_\ell(\epsilon)\left(1+\log
\left[1+\frac{(\alpha\beta)^\ell}{q^4B_\ell(\epsilon)}\right]\right)\right\}\nonumber\\
&\ge&\frac{k}{n}\sum_{i=0}^{n/k-1}\frac{\hat{H}(X_i^\ell,Y_i^\ell)}{\ell}-
\frac{1}{\ell}\log\left\{q^4\left(1+\log\left[1+\frac{(\alpha\beta)^\ell}{q^4}\right]\right)\right\}
-\delta(\epsilon)\nonumber\\
&\ge&\frac{k}{n}\sum_{i=0}^{n/k-1}\rho_{\mbox{\tiny
LZ}}(x_{ik+1}^{ik+k},y_{ik+1}^{ik+k})-\Delta_k'(\ell)-
\frac{1}{\ell}\log\left\{q^4\left(1+\log\left[1+\frac{(\alpha\beta)^\ell}{q^4}\right]\right)\right\}-\delta(\epsilon).
\end{eqnarray}
In the same manner, we can derive a generalized Kraft inequality for the
x-encoder when $y^\ell$ is fixed:
\begin{equation}
\label{kraftx}
\sum_{x^\ell\in\calX^\ell}2^{-\min_{s\in\calS}L[f_{\mbox{\tiny
x}}(s,x^\ell,y^\ell)]}\le
q^2B_\ell(\epsilon)\left(1+\log\left[1+\frac{\alpha^\ell}{q^2B_\ell(\epsilon)}\right]\right),
\end{equation}
and likewise, vice versa:
\begin{equation}
\label{krafty}
\sum_{y^\ell\in\calY^\ell}2^{-\min_{z\in\calZ}L[f_{\mbox{\tiny
y}}(z,y^\ell,x^\ell)]}\le
q^2B_\ell(\epsilon)\left(1+\log\left[1+\frac{\beta^\ell}{q^2B_\ell(\epsilon)}\right]\right).
\end{equation}
Using the same method as above, we arrive at the following individual rate
bounds:
\begin{equation}
\label{Rx}
R_{\mbox{\tiny x}}\ge\frac{k}{n}\sum_{i=0}^{n/k-1}\rho_{\mbox{\tiny
LZ}}(x_{ik+1}^{ik+k}|y_{ik+1}^{ik+k})-\Delta_k'(\ell)-
\frac{1}{\ell}\log\left\{q^2\left(1+\log\left[1+\frac{\alpha^\ell}{q^2}\right]\right)\right\}-\delta(\epsilon),
\end{equation}
and
\begin{equation}
\label{Ry}
R_{\mbox{\tiny y}}\ge\frac{k}{n}\sum_{i=0}^{n/k-1}\rho_{\mbox{\tiny
LZ}}(y_{ik+1}^{ik+k}|x_{ik+1}^{ik+k})-\Delta_k'(\ell)-
\frac{1}{\ell}\log\left\{q^2\left(1+\log\left[1+\frac{\beta^\ell}{q^2}\right]\right)\right\}-\delta(\epsilon).
\end{equation}
Taking the limit superior of $n\to\infty$, followed by the 
limit superior of $k\to\infty$, followed in turn by the limit of $\ell\to\infty$,
and finally, the limit of $\epsilon\downarrow 0$, 
in eqs.\ (\ref{Rx+Ry}), (\ref{Rx}), 
and (\ref{Ry}), we obtain 
\begin{eqnarray}
R_{\mbox{\tiny x}}+R_{\mbox{\tiny
y}}&\ge&\rho(\bx,\by)\\
R_{\mbox{\tiny x}}&\ge&\rho(\bx|\by)\\
R_{\mbox{\tiny y}}&\ge&\rho(\by|\bx).
\end{eqnarray}
This completes the proof of the converse part of Theorem \ref{thm1}.

\section*{Appendix B}
\renewcommand{\theequation}{B.\arabic{equation}}
    \setcounter{equation}{0}
\noindent
{\em Proof of the direct part of Theorem \ref{thm1}.}\\

Consider the partition of the source sequence pair, $(x^n,y^n)$, into
$n/k$ non-overlapping blocks of length $k$,
$(x_{ik+1}^{ik+k},y_{ik+1}^{ik+k})$, $0,1,2,\ldots,n/k-1$.
Select an arbitrarily small $\epsilon_0>0$, and
for each $i$, let us select select a pair of coding rates, $(R_{\mbox{\tiny
x}}^i,R_{\mbox{\tiny y}}^i)$ such that
\begin{eqnarray}
R_{\mbox{\tiny x}}^i&\ge&\rho_{\mbox{\tiny
LZ}}(x_{ik+1}^{ik+k}|y_{ik+1}^{ik+k})+\epsilon_0\\
R_{\mbox{\tiny y}}^i&\ge&\rho_{\mbox{\tiny
LZ}}(y_{ik+1}^{ik+k}|x_{ik+1}^{ik+k})+\epsilon_0\\
R_{\mbox{\tiny x}}^i+R_{\mbox{\tiny y}}^i&\ge&\rho_{\mbox{\tiny
LZ}}(x_{ik+1}^{ik+k},y_{ik+1}^{ik+k})+\epsilon_0,
\end{eqnarray}
so that $R_{\mbox{\tiny x}}=\frac{k}{n}\sum_{i=0}^{n/k-1}R_{\mbox{\tiny
x}}^i$ and
$R_{\mbox{\tiny y}}=\frac{k}{n}\sum_{i=0}^{n/k-1}R_{\mbox{\tiny
x}}^i$ satisfy
\begin{eqnarray}
R_{\mbox{\tiny x}}&\ge&\rho_k(x^n|y^n)
+\epsilon_0\\
R_{\mbox{\tiny y}}&\ge&\rho_k(y^n|x^n)
+\epsilon_0\\
R_{\mbox{\tiny x}}+R_{\mbox{\tiny y}}&\ge&\rho_k(x^n,y^n)
+\epsilon_0,
\end{eqnarray}
and then, in the limit of large $n$ and $k$, $(R_{\mbox{\tiny
x}},R_{\mbox{\tiny y}})$ is in
$R(\bx,\by)$.

Now, for $i=0,1,2,\ldots$, let both $x_{ik+1}^{ik+k}$ and $y_{ik+1}^{ik+k}$ be compressed separately
by random binning encoders with block length $k$,
$\phi_{\mbox{\tiny x}}^i$ and $\phi_{\mbox{\tiny y}}^i$
at rates $R_{\mbox{\tiny x}}^i$ and $R_{\mbox{\tiny y}}^i$, respectively.
This is to say that every $x^k\in\calX^k$ (resp.\ $y^k\in\calY^k$),
is mapped into an index $\phi_{\mbox{\tiny x}}^i(x^k)$ (resp.\
$\phi_{\mbox{\tiny y}}^i(y^k)$) that is selected independently at random across
the range
$\{0,1,2,\ldots,2^{kR_{\mbox{\tiny x}}^i}-1\}$ (resp.\
$\{0,1,2,\ldots,2^{kR_{\mbox{\tiny y}}^i}-1\}$) under the uniform distribution
(independently for every $i$).
Clearly, a block code of length $k$ for $x^k$ (resp.\ $y^k$)
can be implemented using a finite-state machine with no more than $\alpha^k$
(resp.\ $\beta^k$) states.
Note also that this
pair of encoders belongs to the class of slowly time-varying encoders, which are
piece-wise constant along segments of length $k$, as described in Subsection
\ref{formulation}.

Throughout the remaining part of the proof we analyze the probability of error
within each block of length $k$ and show that it tends to zero as
$k\to\infty$, and so, the expected Hamming distance between the decoded
sources and the input sources vanish in the long run. For the sake of
notational simplicity, we henceforth avoid the indexing by $i$. In other
words, with a slight abuse of notation, we replace $x_{ik+1}^{ik+k}$,
$y_{ik+1}^{ik+k}$, $\phi_{\mbox{\tiny x}}^i$, $\phi_{\mbox{\tiny y}}^i$,
$R_{\mbox{\tiny x}}^i$ and $R_{\mbox{\tiny y}}^i$ by
$x^k$, $y^k$, $\phi_{\mbox{\tiny x}}$, $\phi_{\mbox{\tiny y}}$,
$R_{\mbox{\tiny x}}$, and $R_{\mbox{\tiny y}}$, respectively.
Consider now the following decoder that maps the pair $(\phi_{\mbox{\tiny
x}}(x^k),\phi_{\mbox{\tiny y}}(y^k))$ into the decoded estimates of the sources
vectors, $(\hat{x}^k,\hat{y}^k)$:
\begin{equation}
(\hat{x}^k,\hat{y}^k)=
\mbox{arg}\max_{\{(\tx^k,\ty^k):~
\phi_{\mbox{\tiny x}}(\tx^k)=
\phi_{\mbox{\tiny x}}(x^k),~
\phi_{\mbox{\tiny y}}(\ty^k)=
\phi_{\mbox{\tiny y}}(y^k)\}}
u(\tx^k,\ty^k),
\end{equation}
where
\begin{equation}
u(\tx^k,\ty^k)=\min\{[R_{\mbox{\tiny x}}-\rho_{\mbox{\tiny
LZ}}(\tx^k|\ty^k)],~
[R_{\mbox{\tiny y}}-\rho_{\mbox{\tiny LZ}}(\ty^k|\tx^k)],~[R_{\mbox{\tiny x}}+
R_{\mbox{\tiny y}}-\rho_{\mbox{\tiny LZ}}(\tx^k,\ty^k)]\}.
\end{equation}
Observe that $u(x^k,y^k)\ge\epsilon_0$ by (B.1)-(B.3 by (B.1)-(B.3)).
The error event can be presented as the disjoint union of three types of
error events: The first type of error is when $\ty^k=y^k$ and only
$\tx^k\ne\tx^k$, the second type is the other way around,
and the third type is when both $\tx^k\ne x^k$ and $\ty^k\ne y^k$.
Accordingly,
\begin{equation}
P_{\mbox{\tiny e}}(x^k,y^k)= P_{\mbox{\tiny e1}}(x^k,y^k)
+P_{\mbox{\tiny e2}}(x^k,y^k)+
P_{\mbox{\tiny e3}}(x^k,y^k),
\end{equation}
where
\begin{eqnarray}
\label{B10}
P_{\mbox{\tiny e1}}(x^k,y^k)&=&\sum_{\{\tx^k:~u(\tx^k,y^k)\ge
u(x^k,y^k)\}}\mbox{Pr}\{\phi_{\mbox{\tiny x}}(\tx^k)=\phi_{\mbox{\tiny
x}}(x^k)\}\nonumber\\
&\lea&\sum_{\{\tx^k:~kR_{\mbox{\tiny x}}-k\rho_{\mbox{\tiny LZ}}(\tx^k|y^k)\ge
ku(x^k,y^k)\}}2^{-kR_{\mbox{\tiny x}}}\nonumber\\
&\leb&\sum_{\{\tx^k:~LZ(\tx^k|y^k)\le kR_{\mbox{\tiny
x}}-ku(x^k,y^k)+k\hat{\epsilon}(k)
\}}2^{-kR_{\mbox{\tiny x}}}\nonumber\\
&\lec& 2\cdot 2^{kR_{\mbox{\tiny x}}-ku(x^k,y^k)+k\hat{\epsilon}(k)}\cdot
2^{-kR_{\mbox{\tiny x}}}\nonumber\\
&=& 2\cdot 2^{-k[u(x^k,y^k)-\hat{\epsilon}(k)]}\nonumber\\
&\led& 2^{-k[\epsilon_0-\hat{\epsilon}(k)-1/k]},
\end{eqnarray}
where (a) is since $R_{\mbox{\tiny x}}-\rho_{\mbox{\tiny LZ}}(\tx^k|y^k)\ge
u(\tx^k,y^k)$ by
definition of $u(\cdot,\cdot)$, (b) stems from (\ref{conditional-lz}), and (c)
is based on the following consideration: Since $LZ(x^k|y^k)$ is a length
function of a lossless code, the size of the set $\{x^k:~LZ(x^k|y^k)=l\}$
cannot exceed $2^\ell$, and so, for any positive integer $L$,
\begin{equation}
\label{B11}
\bigg|\{x^k: LZ(x^k|y^k)\le L\}\bigg|=
\sum_{l=1}^L\bigg|\{x^k: LZ(x^k|y^k)=l\}\bigg|
\le\sum_{l=1}^L2^l=2^{L+1}-1< 2\cdot 2^L.
\end{equation}
Finally, (d) holds because $u(x^k,y^k)\ge\epsilon_0$, as observed above.
Clearly, $P_{\mbox{\tiny e2}}(x^k,y^k)$ is treated exactly in the same way,
except
that the roles of $x^k$ and $y^k$ are swapped. Therefore,
\begin{equation}
P_{\mbox{\tiny e2}}(x^k,y^k)\le 2^{-k[\epsilon_0-\hat{\epsilon}(k)-1/k]}
\end{equation}
as well. Finally,
\begin{eqnarray}
P_{\mbox{\tiny e3}}(x^k,y^k)&=&\sum_{\{(\tx^k,\ty^k):~u(\tx^k,\ty^k)\ge
u(x^k,y^k)\}}\mbox{Pr}\{\phi_{\mbox{\tiny x}}(\tx^k)=\phi_{\mbox{\tiny
x}}(x^k),~\phi_{\mbox{\tiny y}}(\ty^k)=\phi_{\mbox{\tiny y}}(y^k)\}\nonumber\\
&\le&\sum_{\{(\tx^k,\ty^k):~R_{\mbox{\tiny x}}+R_{\mbox{\tiny
y}}-\rho_{\mbox{\tiny LZ}}(\tx^k,\ty^k)\ge
u(x^k,y^k)\}}2^{-k(R_{\mbox{\tiny x}}+R_{\mbox{\tiny y}})}\nonumber\\
&=&\sum_{\{(\tx^k,\ty^k):~LZ(\tx^k,\ty^k)\le
k(R_{\mbox{\tiny x}}+R_{\mbox{\tiny y}})-ku(x^k,y^k)+k\epsilon(k)
\}}2^{-k(R_{\mbox{\tiny x}}+R_{\mbox{\tiny y}})}\nonumber\\
&<& 2\cdot 2^{k(R_{\mbox{\tiny x}}+R_{\mbox{\tiny
y}})-ku(x^k,y^k)+k\epsilon(k)}\cdot
2^{-k(R_{\mbox{\tiny x}}+R_{\mbox{\tiny y}})}\nonumber\\
&\le& 2^{-k[\epsilon_0-\epsilon(k)-1/k]},
\end{eqnarray}
and so, overall,
$P_{\mbox{\tiny e}}(x^k,y^k)$ tends to zero as $k\to\infty$.
This completes the proof of the direct part of Theorem \ref{thm1}.

\section*{Appendix C}
\renewcommand{\theequation}{C.\arabic{equation}}
    \setcounter{equation}{0}

Since the inequality $\rho^+(\bx,\by)\ge \rho^-(\bx,\by)$ is obvious, it
is enough to prove
the reverse inequality,
$\rho^+(\bx,\by)\le \rho^-(\bx,\by)$.
For a given $n$, consider the following encoding of $(x^n,y^n)$ in blocks of
length $k$, where $k$ is assumed to divide $n$. For each
$k$-block, $(x_{ik+1}^{ik+k},y_{ik+1}^{ik+k})$, $i=0,1,\ldots,n/k-1$, the encoder compares the
length functions of three compression schemes:
\begin{enumerate}
\item Scheme A applies LZ compression of the sequence of pairs
$(x_{ik+1},y_{ik+1}),\ldots
(x_{ik+k},y_{ik+k})$ using $LZ(x_{ik+1}^{ik+k},y_{ik+1}^{ik+k})$ bits.
\item Scheme B first compresses $x_{ik+1}^{ik+k}$ to $LZ(x_{ik+1}^{ik+k})$
bits and then compresses $y_{ik+1}^{ik+k}$ into
$LZ(y_{ik+1}^{ik+k}|x_{ik+1}^{ik+k})$ bits by utilizing
$x_{ik+1}^{ik+k}$ as side information available at both ends.
\item Scheme C is the same as Scheme B, except that the roles of
$x_{ik+1}^{ik+k}$ and $y_{ik+1}^{ik+k}$ are interchanged.
\end{enumerate}
The encoder selects the shortest code among those of schemes A, B and C, and
adds a header of two
flag bits to indicate to the decoder which one of the three schemes was chosen. The overall
coding rate is therefore
\begin{equation}
\frac{1}{n}\sum_{i=0}^{n/k-1}[k\rho_{\mbox{\tiny LZ}}^-(x_{ik+1}^{ik+k},y_{ik+1}^{ik+k})+2]=
\frac{k}{n}\sum_{i=0}^{n/k-1}\rho_{\mbox{\tiny LZ}}^-(x_{ik+1}^{ik+k},y_{ik+1}^{ik+k})+\frac{2}{k}.
\end{equation}
This is a block code of length $k$, and as such, it can be implemented by a finite-state machine with no more than
\begin{equation}
s=\sum_{i=0}^{k-1}\alpha^i\beta^i=\frac{\alpha^k\beta^k-1}{\alpha\beta-1}<\alpha^k\beta^k 
\end{equation}
states (see footnote no.\ 1). Therefore,
\begin{equation}
\varrho_{\alpha^k\beta^k}(x^n,y^n)\le
\frac{k}{n}\sum_{i=0}^{n/k-1}\rho_{\mbox{\tiny LZ}}^-(x_{ik+1}^{ik+k},y_{ik+1}^{ik+k})+\frac{2}{k},
\end{equation}
where $\varrho_s(x^n,y^n)$ and its corresponding limits were defined after Theorem \ref{thm2}.
Consequently,
\begin{equation}
\varrho_{\alpha^k\beta^k}(\bx,\by)=\limsup_{n\to\infty}
\varrho_{\alpha^k\beta^k}(x^n,y^n)\le
\limsup_{n\to\infty}\frac{k}{n}\sum_{i=0}^{n/k-1}\rho_{\mbox{\tiny LZ}}^-(x_{ik+1}^{ik+k},y_{ik+1}^{ik+k})+\frac{2}{k}.
\end{equation}
Upon taking the limit superior of $k\to\infty$, we get
\begin{equation}
\label{direct}
\varrho_\infty(\bx,\by)=\limsup_{k\to\infty}\varrho_{\alpha^k\beta^k}(\bx,\by)
=\lim_{k\to\infty}\varrho_{\alpha^k\beta^k}(\bx,\by)
\le \rho^-(\bx,\by).
\end{equation}

On the other hand, as a lower bound to the lossless compression ratio
for the $i$th $k$-block ($i=0,1,\ldots,n/k-1)$) by an $s$-state encoder, we can
apply the second to the last line of eq.\ (\ref{Rx+Ry}) with $\epsilon=0$ to obtain:
\begin{eqnarray}
\frac{L_i}{k}&\ge&
\frac{\hH_\ell(X_i^\ell,Y_i^\ell)}{\ell}-\delta_s(\ell)\nonumber\\
&\ge&\rho_{\mbox{\tiny
LZ}}(x_{ik+1}^{ik+k},y_{ik+1}^{ik+k})-\delta_s(\ell)-\Delta_k'(\ell)
\end{eqnarray}
where $\ell$ divides $k$,
\begin{equation}
\delta_s(\ell)=
\frac{1}{\ell}\log\left\{s^2\left[1+\log\left(1+\frac{\alpha^\ell\beta^\ell}{s^2}\right)\right]\right\}
\end{equation}
and $\Delta_k'(\ell)$ is as defined in Subsection \ref{background}.
But $\hH_\ell(X_i^\ell,Y_i^\ell)/\ell$ can be decomposed also as
\begin{equation}
\frac{\hH(X_i^\ell)}{\ell}+\frac{\hH(Y_i^\ell|X_i^\ell)}{\ell}\ge
\rho_{\mbox{\tiny LZ}}(x_{ik+1}^{ik+k})+\rho_{\mbox{\tiny LZ}}(y_{ik+1}^{ik+k}|x_{ik+1}^{ik+k})-2\Delta_k'(\ell)
\end{equation}
as well as
\begin{equation}
\frac{\hH(Y_i^\ell)}{\ell}+\frac{\hH(X_i^\ell|Y_i^\ell)}{\ell}\ge
\rho_{\mbox{\tiny LZ}}(y_{ik+1}^{ik+k})+\rho_{\mbox{\tiny
LZ}}(x_{ik+1}^{ik+k}|y_{ik+1}^{ik+k})-2\Delta_k'(\ell),
\end{equation}
which together imply that
\begin{eqnarray}
\frac{L_i}{k}&\ge&
\max\{\rho_{\mbox{\tiny
LZ}}(x_{ik+1}^{ik+k},y_{ik+1}^{ik+k}),\rho_{\mbox{\tiny
LZ}}(x_{ik+1}^{ik+k})+\rho_{\mbox{\tiny LZ}}(y_{ik+1}^{ik+k}|x_{ik+1}^{ik+k}),\nonumber\\
& &\rho_{\mbox{\tiny
LZ}}(y_{ik+1}^{ik+k})+\rho_{\mbox{\tiny LZ}}(x_{ik+1}^{ik+k}|y_{ik+1}^{ik+k})\}-\delta_s(\ell)-2\Delta_k'(\ell)\nonumber\\
&=&\rho_{\mbox{\tiny
LZ}}^+(x_{ik+1}^{ik+k},y_{ik+1}^{ik+k})-\delta_s(\ell)-2\Delta_k'(\ell).
\end{eqnarray}
Thus, the $s$-state compressibility of $(x^n,y^n)$ is lower bounded as
follows:
\begin{equation}
\varrho_s(x^n,y^n)\ge\frac{k}{n}\sum_{i=0}^{n/k-1}\rho_{\mbox{\tiny LZ}}^+(x_{ik+1}^{ik+k},y_{ik+1}^{ik+k})-
\delta_s(\ell)-2\Delta_k'(\ell),
\end{equation}
which leads to
\begin{equation}
\varrho_s(\bx,\by)=\limsup_{n\to\infty}\varrho_s(x^n,y^n)\ge
\limsup_{n\to\infty}\frac{k}{n}\sum_{i=0}^{n/k-1}\rho_{\mbox{\tiny
LZ}}^+(x_{ik+1}^{ik+k},y_{ik+1}^{ik+k})-
\delta_s(\ell)-2\Delta_k'(\ell).
\end{equation}
Upon taking $k$ (and then $\ell$) to infinity (yet keeping $s$ fixed),
we obtain
\begin{equation}
\varrho_s(\bx,\by)\ge\rho^+(\bx,\by),
\end{equation}
and so,
\begin{equation}
\varrho_\infty(\bx,\by)=\lim_{s\to\infty}\varrho_s(\bx,\by)\ge \rho^+(\bx,\by),
\end{equation}
which together with (\ref{direct}),
yields
\begin{equation}
\rho^+(\bx,\by)\le \rho^-(\bx,\by).
\end{equation}
Consequently, $\rho^+(\bx,\by)=\rho^-(\bx,\by)=\rho(\bx,\by)$, and the proof of
Theorem \ref{thm2} is complete.


\begin{thebibliography}{AA}

\bibitem{CK11}
I.~Csisz\'ar and J.~K\"orner, {\em Information Theory -- Coding Theorems for
Discrete Memoryless Systems}, Cambridge University Press, New York 2011.

\bibitem{Draper04}
S.~C.~Draper, ``Universal incremental Slepian--Wolf coding,''
{\it Proc.\ Annual Allerton Conference on Communication, Control, and
Computing}, Monticello, IL,
October 2004.

\bibitem{Kolmogorov68}
A.~N.~Kolmogorov, ``Logical basis for information theory and probability
theory,'' {\em IEEE Trans.\ Inform.\ Theory}, vol.\ IT-14, no.\ 5, pp.\
662-664, September 1968.

\bibitem{me00}
N.~Merhav, ``Universal detection of messages via finite--state channels,''
{\em IEEE Trans.\ Inform.\ Theory},
vol.\ 46, no.\ 6, pp.\ 2242--2246, September 2000.

\bibitem{me16}
N.~Merhav, ``Universal decoding for source-channel coding with side
information,'' {\em Communications in Information and Systems}, vol.\ 16, no.\
1, pp.\ 17-58, 2016.

\bibitem{me23}
N.~Merhav, ``A universal ensemble for sample-wise lossy compression,''
{\em Entropy}, 2023, 25(8), 1199;
{\tt https://doi.org/10.3390/e25081199},
August 2023.

\bibitem{Shulman03}
N.~Shulman, {\em Communication over an Unknown Channel in Common
Broadcasting}, Ph.D.\ thesis, Department of Electrical Engineering -- Systems, Tel Aviv University, 2003.

\bibitem{SF02}
N.~Shulman and M.~Feder, ``Source broadcasting with an unknown amount of
receiver side information,'' {\em Proc.\ 2002 IEEE Information Theory
Workshop}, Bangalore, India, pp.\ 127--130, October 2002.

\bibitem{SW73}
D.~Slepian and J.~K.~Wolf, ``Noiseless coding of correlated information
sources,'' {\em IEEE Trans.\ Inform.\ Theory}, vol.\ IT-19, pp.\ 471--480,
July 1973.

\bibitem{UK03}
T.~Uyematsu and S.~Kuzuoka, ``Conditional Lempel-Ziv complexity and its
application to source coding theorem with side information,''
{\em IEICE Trans.\ Fundamentals}, Vol.\ E86-A, no.\ 10, pp.\ 2615--2617,
October 2003.

\bibitem{Ziv84}
J.~Ziv, ``Fixed-rate encoding of individual sequences with side information'',
{\em IEEE Transactions on Information Theory\/},
vol.~IT--30, no.~2, pp.~348--452, March  1984.

\bibitem{Ziv85}
J.~Ziv, ``Universal decoding for finite-state channels,'' 
{\em IEEE Trans.~Inform.~Theory\/},
vol.~IT--31, no.~4, pp.~453--460, July 1985.

\bibitem{ZL78}
J.~Ziv and A.~Lempel, ``Compression of individual sequences via 
variable-rate coding,''
{\em IEEE Trans.~Inform.~Theory\/},
vol.~IT--24, no.~5, pp.~530--536, September 1978.

\bibitem{ZL70}
A.~K.~Zvonkin and L.~A.~Levin, ``The complexity of finite objects and
the development of the concepts of information and randomness by means of the
theory of algorithms,'' {\em Russian Mathematical Surveys},
vol.\ 25, no.\ 6, pp.\ 83--124, 1970.
\end{thebibliography}
\end{document}